# AI for Beyond 5G Networks: A Cyber-Security Defense or Offense Enabler?

Chafika Benzaïd* and Tarik Taleb†
*†Aalto University, Espoo, Finland
†University of Oulu, Oulu, Finland
†Sejong University, Seoul, South Korea
Email: *chafika.benzaid@aalto.fi, †tarik.taleb@aalto.fi

*Abstract*—Artificial Intelligence (AI) is envisioned to play a pivotal role in empowering intelligent, adaptive and autonomous security management in 5G and beyond networks, thanks to its potential to uncover hidden patterns from a large set of time-varying multi-dimensional data, and deliver faster and accurate decisions. Unfortunately, AI's capabilities and vulnerabilities make it a double-edged sword that may jeopardize the security of future networks. This paper sheds light on how AI may impact the security of 5G and its successive from its posture of defender, offender or victim, and recommends potential defenses to safeguard from malevolent AI while pointing out their limitations and adoption challenges.

*Index Terms*—5G, B5G, AI, ML, Security

## I. Introduction

5G and beyond networks hold out the promise of delivering ultra-low latency, ultra-high throughput, ultra-high reliability, ultra-low energy usage, and massive connectivity. Achieving these promises will pave the way to a new breed of applications, including autonomous driving, industry 4.0, augmented and virtual reality, collaborative gaming, near real-time remote surgery, and teleportation. However, the diversity of services/applications and the growing number of connected things envisaged in the networks of tomorrow will open up new and increasingly broad cyber threats, posing security and privacy risks [1]. Thus, it is imperative to build up effective and sustainable security measures that can deal with the ever-evolving threat landscape and security requirements in 5G and its successive in order to fully reap their benefits.

Considering the increasing number of vulnerabilities, the growing sophistication of cyber threats, the high traffic volume, and the diverse technologies (e.g., SDN, NFV) and services that will shape the next-generation wireless networks, the reliance on traditional security management approaches may no longer suffice and need to be rethought to cope with this challenging environment. A promising direction is the adoption of Artificial Intelligence (AI) to empower intelligent, adaptive and autonomous security management, allowing timely and cost-effective detection and mitigation of security threats. Indeed, AI has the potential of uncovering hidden patterns from a large set of time-varying multi-dimensional data, and delivering faster and accurate decisions. In response to the trend of integrating AI, particularly Machine Learning (ML), into telecommunication networks, the ITU-T Focus Group[1] on Machine Learning for Future Networks including 5G (FG-ML5G) has recently released a unified architectural framework for ML in future networks.

Though the key role of AI in enforcing security in 5G and beyond networks is incontestable, its capabilities make it a double-edged sword. Indeed, AI's capabilities, shored up by the envisioned ultra-high bandwidth and the massive proliferation of connected devices, will usher in a new era of sophisticated cyber-attacks that are autonomous, scalable, stealthy and faster. Moreover, the major role that AI systems will play in empowering self-managing functionalities (e.g., self-optimization, self-healing, and self-protecting) in future networks makes AI an attractive target for cyber-criminals. An adversary may leverage the vulnerabilities of AI systems to subvert their performance and security. Recognizing the seriousness of AI's dangers, ETSI has launched a new Industry Specification Group on Securing Artificial Intelligence (ISG SAI)[2]. The purpose of ISG SAI is to develop technical specifications to mitigate threats stemming from the deployment of AI in ICT field.

This paper aims to shed light on how AI may impact the security of 5G and beyond networks from its posture of defender, offender or victim (See Fig. 1). The rest of this paper is organized as follows. Section II discusses some prospective applications of AI to bolster the security of future networks. Section III explores the potential risks arising from AI systems exploited either as an instrument or as a target to impede the security of 5G and beyond networks. Section IV surveys possible defense measures that could be adopted to safeguard from malevolent AI, and recommends where those measures could be enforced into the FG-ML5G unified architecture. Finally, the paper concludes in Section V.

## II. AI's Potential for Cyber-Security in B5G Networks

5G and beyond networks will be characterized by massive number of connected devices, high traffic volume, and diverse technologies (e.g., SDN, NFV) and services, leading to a complex and dynamic cyber-threat landscape. A promising direction to deal with this challenging threat landscape is the adoption of AI, thanks to its potential in empowering intelligent, adaptive and autonomous security management. In what follows, some prospective applications of AI for security in future networks are discussed.

---

[1]https://www.itu.int/en/ITU-T/focusgroups/ml5g/Pages/default.aspx

[2]https://www.etsi.org/committee/1640-sai

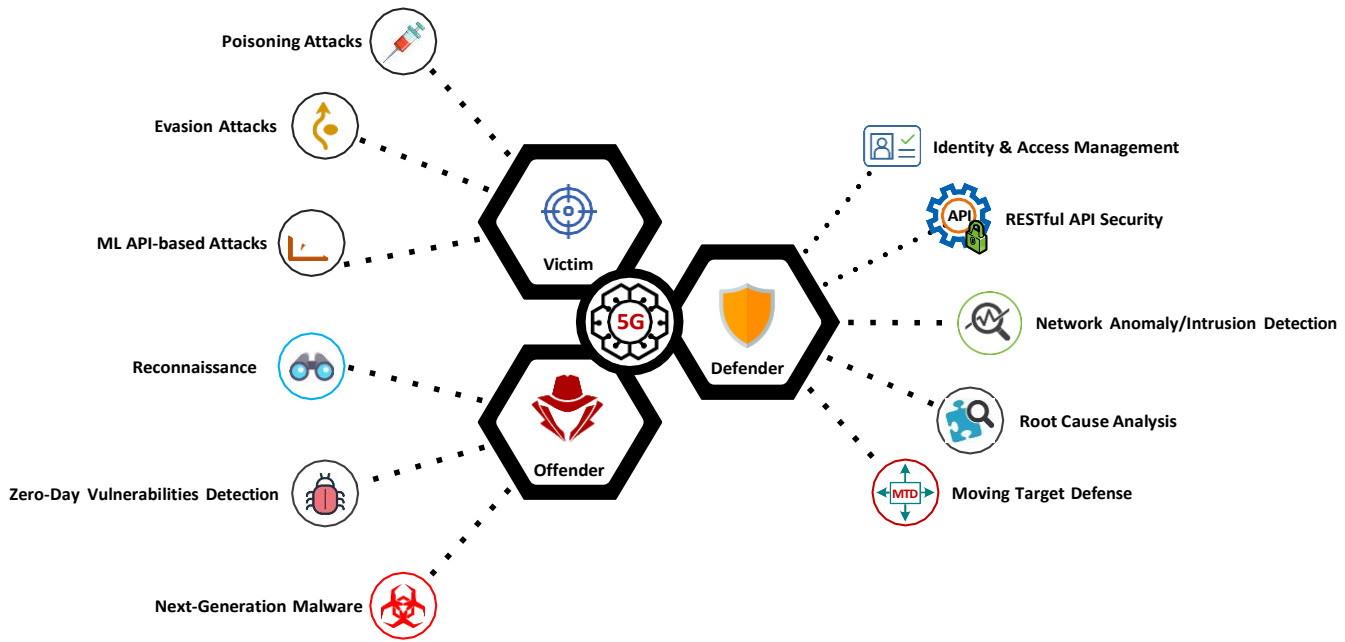

Fig. 1: AI's Impact on Security in 5G and Beyond Networks from its Posture of Defender, Offender or Victim.

*A. Identity and Access Management*

Authentication and authorization services play a key role in 5G and beyond security, preventing impersonation attacks and controlling access privileges of involved entities (physical or virtual). However, the envisaged massive machine type communications (mMTC) and ultra reliable and low-latency communications (URLLC) use cases require the support of extreme device density, energy efficiency and low-latency capabilities. Meanwhile, the adoption of small cell densification in 5G networks will induce frequent handovers and consequently frequent authentications, resulting in increased latency. Thus, efficient, scalable and fast authentication mechanisms are essential to cope with the aforementioned stringent requirements. AI is deemed to play a pivotal role in achieving this goal. In fact, emerging authentication and authorization schemes are increasingly relying on multiple non-cryptographic attributes, associated with users, resources and environment (e.g., time and location), to determine the identity and authorizations of a given entity. The merit of AI stems from its ability to automatically combine these diverse and time-varying attributes to provide continuous authentication and dynamically enforce fine-grained access policies. Fang *et al.* [2] introduced ML-based intelligent authentication approaches by opportunistically leveraging physical layer attributes (e.g., carrier frequency offset, channel impulse response, and receiving signal strength indication) to achieve continuous and situation-aware authentication in 5G and beyond networks. The work in [3] proposes a holistic authentication and authorization approach relying on online ML and trust management for achieving adaptive access control in a large-scale and dynamic IoT environment. The proposed access control scheme intelligently exploits the time-varying features of the transmitter, i.e, communication-related, hardware-related attributes and user behaviors, to refine and update access policies on run-time.

*B. RESTful API Security*

Application Programming Interfaces (REST APIs) play a vital role in the 5G ecosystem as they empower service exposure across different networks. This is why 3GPP decided that the implementation of both northbound interfaces (NBIs) and service based interfaces (SBIs) should be based on RESTful APIs. Due to their importance for 5G and beyond networks, APIs will most probably become a primary target for attackers. Indeed, the recent ENISA's "threat landscape for 5G networks" report[3] has identified API exploitation/abuse as a nefarious threat against 5G assets, resulting in information leakage/alteration/destruction, identity theft as well as service unavailability. Thus, API security is a cornerstone to protect 5G and beyond networks. However, the wide variety of APIs and the sheer volume of API traffic envisioned in the next-generation mobile networks make the identification and mitigation of API threats a complex task. AI-driven API security is the new trend to cope with the aforementioned challenges. In fact, AI has the capability of uncovering patterns in vast amounts of multidimensional data, allowing continuous and proactive monitoring and detection of API attacks and fostering their automatic mitigation.

*C. Network Anomaly/Intrusion Detection and Prediction*

To fulfill the stringent reliability and availability requirements of 5G and beyond networks, a timely detection and prediction of anomalous behaviors due to malicious or accidental actions is paramount. Indeed, the early identification of potential problems in the network enables fast reaction to them, preventing extreme malicious damage, service degradation and financial loss [4]. An anomaly refers to "a pattern that does not conform to expected normal behavior" [5].

---
[3]https://www.enisa.europa.eu/publications/enisa-threat-landscape-for-5g-networks

ETSI ENI (Experiential Network Intelligence) ISG (Industry Specification Group) [6] has identified AI usage as a requirement to recognize abnormal traffic patterns that can lead to service unavailability or security threats in next-generation networks. The AI has proven its ability in uncovering hidden patterns from a large set of time-varying multi-dimensional data. The work in [7] proposes an anomaly detection solution for the self-healing of Radio Access Networks (RAN) in 5G networks, where the anomaly patterns are identified leveraging the clustering algorithm DBSCAN. The authors demonstrated the effectiveness of their solution in detecting anomalies caused by radio attenuation and SDN misconfiguration. The authors in [8] proposed a Deep Learning (DL)-based solution for detecting anomalies due to cell outages and congestion. The use of shallow and deep learning approaches for detecting and forecasting network intrusions has attracted considerable attention [9]. Krayani *et al.* [10] devised a Dynamic Bayesian Network (DBN) model to detect jamming attack in Orthogonal Frequency Division Multiplexing (OFDM)-based cognitive radio networks.

*D. Root Cause Analysis*

Once an anomaly alarm is triggered, its underlying cause needs to be determined. In fact, an accurate identification of the cause of faults and security incidents is a key to empower self-organizing system, establish effective mitigation strategies, perform network forensics and even assign liability. However, given the complexity and heterogeneity of emerging mobile networks, coupled with the increasing number of Key Performance Indicators (KPIs) and data related to end-users, services and networks, the root cause diagnosis becomes highly intractable. Thus, a manual assessment of root causes based on expert knowledge is complex and time- and effort-consuming task. AI has been recognized as an appealing option for fostering self root cause analysis, thanks to its ability to process a large amount of data, uncover complex non-linear relationships within the data, and deliver faster and accurate decisions. For instance, Zhang *et al.* [11] proposed a DL-based root cause analysis of faults in a cellular RAN, leveraging both supervised classification (Auto-Encoder + Decision Tree) and unsupervised clustering (Auto-Encoder + agglomerative clustering). In the smart manufacturing vertical domain, AI-driven root cause analysis can not only help in tracing the root of failure events, but also in predicting future anomalies, leading to improved operational efficiency and reduced unplanned downtime.

*E. Moving Target Defense*

The static nature of network and service configurations once deployed facilitates the adversary mission in exploring and exploiting the unchanging vulnerability surface. In fact, the vulnerability persistence gives the attacker the advantage of time to understand the attack surface and choose the best-fitting attack technique. The Moving Target Defense (MTD) has emerged as an effective proactive security solution to address this problem. Indeed, NIST [12] has recognized MTD as an enhanced security requirement for system and communications protection. MTD approaches aim at increasing the attacker's effort and cost by dynamically and constantly changing the attack surface over runtime. The MTD can be established through various implementations including, IP address shuffling, VM migration, network path diversification, and replication of software or network resources. The flexibility and dynamicity opportunities provided by virtualization (i.e., Network Function Virtualization) and programmability (i.e., Software Defined Networking) will foster the implementation of MTD mechanisms in 5G and beyond networks, leading to more resilient networks. The MTD paradigm is an appealing security strategy for various vertical application domains, such as IoT and automotive domains. For example, reconnaissance, impersonation and DoS attacks against in-vehicle networks can be prevented by adopting a dynamic address/ID shuffling strategy. Meanwhile, path diversification and topology shuffling can be used for improving resilience of Inter-vehicles wireless communications to eavesdropping and jamming attacks. However, it is worth stating that the security benefits of MTD come at the expense of reconfiguration cost and/or service unavailability. Thus, achieving the desired balance between the security effectiveness of MTD and the induced cost is of utmost importance. AI techniques, including game theory, genetic algorithms and ML, have been considered highly relevant to devise smart MTD mechanisms that can intelligently decide changes to make on the network and service configuration in order to meet the security/performance trade-off [13]. For instance, Albanese et al. [14] used a Reinforcement Learning (RL) model to develop a MTD strategy to resist stealthy botnets by periodically altering the placement of detectors.

III. AI's Threats Against B5G Security

In view of the major role AI systems will play in 5G and beyond networks, their security risks represent a key aspect to consider. In fact, the potential threats emanating from the use of AI systems can be broadly classified into two categories, namely: (i) *AI as an instrument* to build sophisticated cyber attacks leveraging the capabilities of AI; and (ii) *AI as a target* where the vulnerabilities of AI systems are exploited to undermine their performance and security.

*A. AI as an Instrument*

The AI capability to learn and adapt will pave the way for a new era of AI-powered cyber-attacks that are autonomous, scalable, stealthy and faster. Combining AI's capabilities with the envisioned ultra-high bandwidth and the massive proliferation of connected devices, 5G and beyond networks will doubtlessly see a wide use of AI-driven cyber-attacks. Attackers can utilize AI to conduct a rapid and efficient reconnaissance of the target network in order to identify, for instance, devices deployed, operating systems and services used, ports open, and accounts, especially those with admin privileges. The insights gathered from the reconnaissance phase can be leveraged by AI to learn and prioritize vulnerabilities that may be exploited to launch a large-scale network attack. For instance, an AI-based botnet can automatically identify zero-day vulnerabilities in IoT devices and exploit them to perform a large-scale distributed denial of service attack by creating a signaling storm against the 5G RAN resources. AI is also expected to drive the development of next-generation malware that are able to operate autonomously. Autonomous malware will hold the ability to observe the environment, smartly select its target and the most effective lateral movement technique to reach it without raising suspicion. DeepLocker is a proof-of-concept of autonomous malware developed by IBM, which uses a deep neural network model to trigger the malicious payload if the intended victim is reached. The victim is identified through a set of attributes, including geo-location, user activity, and environment features. Another potential offensive use of AI is to conduct an identity spoofing attack by learning and mimicking the behavior of legitimate entities.

### B. AI as a Target

5G and beyond networks will be heavily reliant on AI to enable fully autonomous management capabilities (e.g., self-configuration, self-optimization, self-healing, and self-protecting) [15], making AI an attractive target for attackers. In fact, AI systems, particularly ML systems, can be influenced to learn wrong models, make erroneous decisions/predictions, or leak confidential information. The attacks against ML systems are considered *causative* if they target the training phase or *exploratory* if they aim at the inference phase. They can be conducted in a *white-box*, *gray-box* or *black-box* setting, depending on whether the attacker has, respectively, full, partial or no knowledge about the training data, the learning algorithm and its hyper-parameters. The adversary may perform *indiscriminate* attacks to cause the misclassification of any sample or *targeted* attacks to lead to misclassification of a specific sample. By attacking a ML system, the adversary may decide to break its *integrity* by evading detection without affecting normal behavior of the system; its *availability* by deteriorating the system usability; or its *privacy* by gaining sensitive information about the training data, the ML system or its users. In what follows, we summarize the main attacks against ML systems.

*1) Poisoning Attacks:* In poisoning attacks, also referred to as causative attacks, an attacker aims at influencing the learning outcome to his advantage by tampering with data or the learning algorithm at training phase. The appeal of this attack stems from the constant retraining requirement of a learning model to account for the new data distribution, giving the attackers the opportunity to poison the trained model. The poisoning attack can be mounted using different strategies, namely:

- *Data Injection:* This strategy is used when the attacker has no access to the training data. It aims at altering the data distribution by feeding carefully crafted malicious samples into the training dataset while keeping original samples unchanged.
- *Data Manipulation:* The attacker is assumed to have a full access to the training data, allowing him to directly contaminate the original data used for training the learning model. The contamination can be performed by either flipping labels (e.g., benign to malicious and vice-versa) or introducing small perturbations on input features.
- *Logic Corruption:* The attacker focuses on interfering with the learning algorithm or its learning logic. This strategy can be used against models that leverage distributed learning (e.g., federated learning), which relies on several agents for training. Thus, a malicious agent may manipulate the local model parameters to compromise the global model.

An illustrative example of how a poisoning attack works is given in Fig. 2. Let consider a cognitive radio transceiver which performs a real-time spectrum sensing and determines idle channels for transmission using ML techniques. An adversary can pollute the spectrum sensing data, used for retraining the ML model, by transmitting for a short period when the channel is idle. Thus, the poisoned model is fooled into making the wrong decision of not transmitting when the channel is unoccupied.

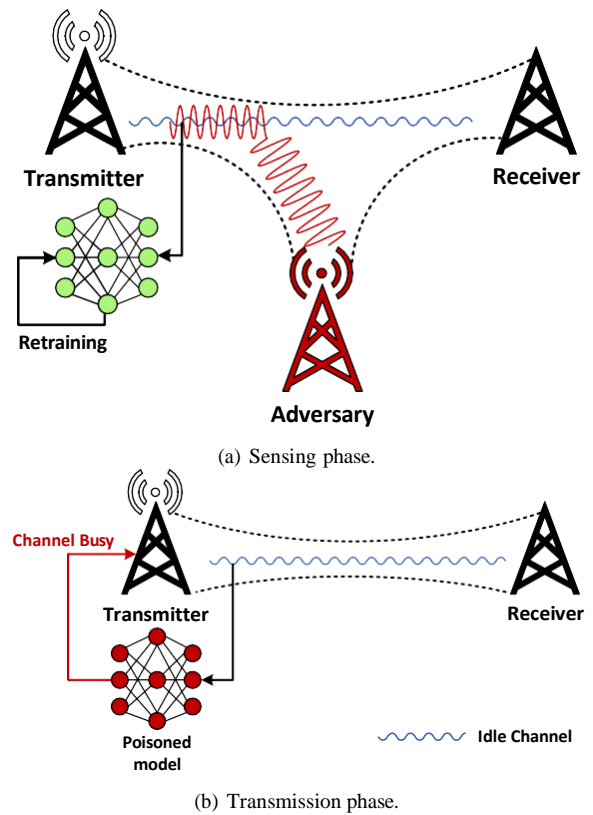

(a) Sensing phase.

(b) Transmission phase.

Fig. 2: Spectrum Data Poisoning in Cognitive Radio Networks.

*2) Evasion Attacks:* An evasion attack targets the inference stage. Unlike poisoning attacks, these attacks require no influence over the training process. The attacker seeks to escape the learned model at test time by introducing small perturbations to the input instances. Such perturbations are called adversarial examples.

Fig. 3 illustrates an exemplary evasion attack against a ML model trained to authenticate IoT devices requesting access to Multi-access Edge Computing (MEC) services. The model leverages the unique features of the physical layer (e.g., carrier frequency offset and receiving signal strength indication) to distinguish between legitimate devices and illegal devices. A smart illegitimate device can fool the trained model to wrongly identify it as a legal device by generating spoofed wireless signals mimicking the signals of the legal device.

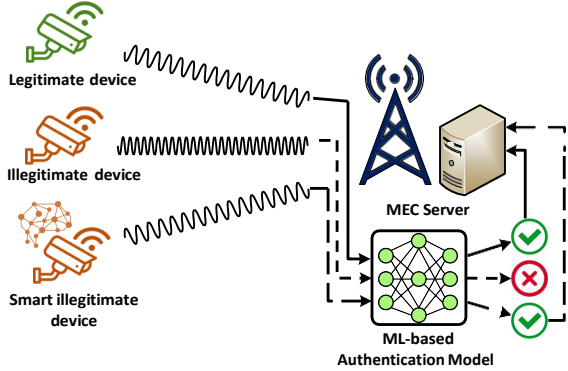

Fig. 3: Adversarial Identity Spoofing.

*3) Model's API-based Attacks:* The emergence of ML-as-a-Service (MLaaS) paradigm makes ML models susceptible to new attacks, namely: model inversion attack, model extraction attack, and membership inference attack. The *model inversion* attack aims to recover the training data by leveraging the outputs of the targeted ML model. Meanwhile, the *model extraction* attack focuses on revealing the model's architecture and parameters to reproduce a (near)-equivalent ML model, by observing the model's predictions and/or execution time. The purpose of a *membership inference* attack is to determine whether a sample has been used to train the target ML model, by exploiting the model's output.

Fig. 4 shows a potential scenario where model extraction attack is conducted to facilitate a subsequent evasion attack against a network anomaly detection module. The module leverages an online ML service to detect suspicious activities in the RAN, such as the symptoms of a signaling DDoS attack. An adversary, with only access to the inference API of the target ML model, can uncover its architecture and parameters to build a surrogate model approximating the target model. The substitute model is then integrated into a botnet malware aiming to generate malicious traffic (e.g., a signaling storm) that can fly under the radar of the anomaly detection module; that is, it will be identified as benign traffic.

*C. Mapping of Adversarial Attacks to ITU-T's Unified Architecture for ML in 5G and Beyond Networks*

The ITU-T's ML5G focus group has recently proposed a unified architecture for ML in 5G and future networks. As depicted in Fig. 5, the unified architecture comprises the following components:

- *ML pipeline:* It is a logical representation of a ML-based network application. The ML pipeline consists of: (1) a **source node (src)** which generates the raw data to feed into the ML model; (2) a **collector (C)** which collects the data from the src; (3) a **preprocessor node (PP)** which prepares data to fit for ML model by performing different data processing operations, encompassing data cleansing, transformation and aggregation; (4) a **model node (M)** which represents a ML model; (5) a **policy node (P)** which leverages the output of M to apply the suitable

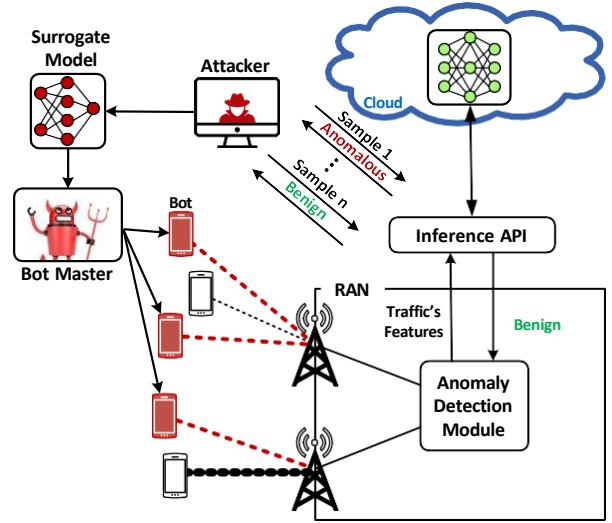

Fig. 4: Model Extraction for Subsequent Evasion Attack Against an AI-based Network Anomaly Detection Module.

   policy depending on the considered use case; (6) a **sink node** on which the selected policy takes action; and (7) a **distributor node (D)** which is in charge of identifying the sinks and distributing policies to the corresponding sinks.
- *ML Sandbox:* It is an isolated domain which serves to train, test and evaluate ML models before deploying them into production. To this end, the sandbox can use synthetic data generated by a simulator and/or real data collected from the network.
- *ML Function Orchestrator (MLFO):* It manages and orchestrates the ML pipeline life-cycle based on ML intent and/or dynamic network conditions. MLFO's responsibilities include the placement of ML pipelines, the flexible chaining of the ML pipeline components to adapt to the underlay network dynamics, the monitoring of the ML model performance, and the selection/reselection of a ML model based on its performance.

Fig. 5 illustrates the mapping of the adversarial ML attacks (i.e., poisoning attacks, evasion attacks and API-based attacks) to the components of the ML5G unified architecture.

IV. POTENTIAL DEFENSE MECHANISMS

This section introduces the potential defense mechanisms that could be adopted to increase resilience to threats targeting AI systems.

*A. Adversarial Machine Learning*

Adversarial Machine Learning (AML) [16] aims at improving the robustness of ML techniques to adversarial attacks by assessing their vulnerabilities and devising appropriate defense measures.

*1) Defenses Against Poisoning Attacks:* Several countermeasures have been proposed against poisoning attacks, which can be broadly categorized into input validation and robust learning. *Input validation* seeks to sanitize the (re)training data from malicious and abnormal samples before feeding it into the ML model. Outlier detection is a common defensive technique used to identify and remove suspicious samples from the training dataset. However, this technique can be bypassed by crafting poisoned samples that can mislead the learning process while remaining within the genuine data distribution. Reject On Negative Impact (RONI) approach sanitizes data by removing samples that have a detrimental impact on the learning performance. Micromodels strategy performs data cleaning by first generating multiple micro-models trained on disjoint subset of input samples. The micro-models are then combined in a majority voting scheme to eliminate the anomalous training data subsets. Clustering-based techniques have been used to mitigate the label flipping attack. These techniques consist in dividing the training data into clusters, where the samples within the same cluster are relabeled using the most common label in this cluster. Unlike input validation, *robust learning* aims at developing learning algorithms that are robust to training data contamination by leveraging robust statistics techniques [17].

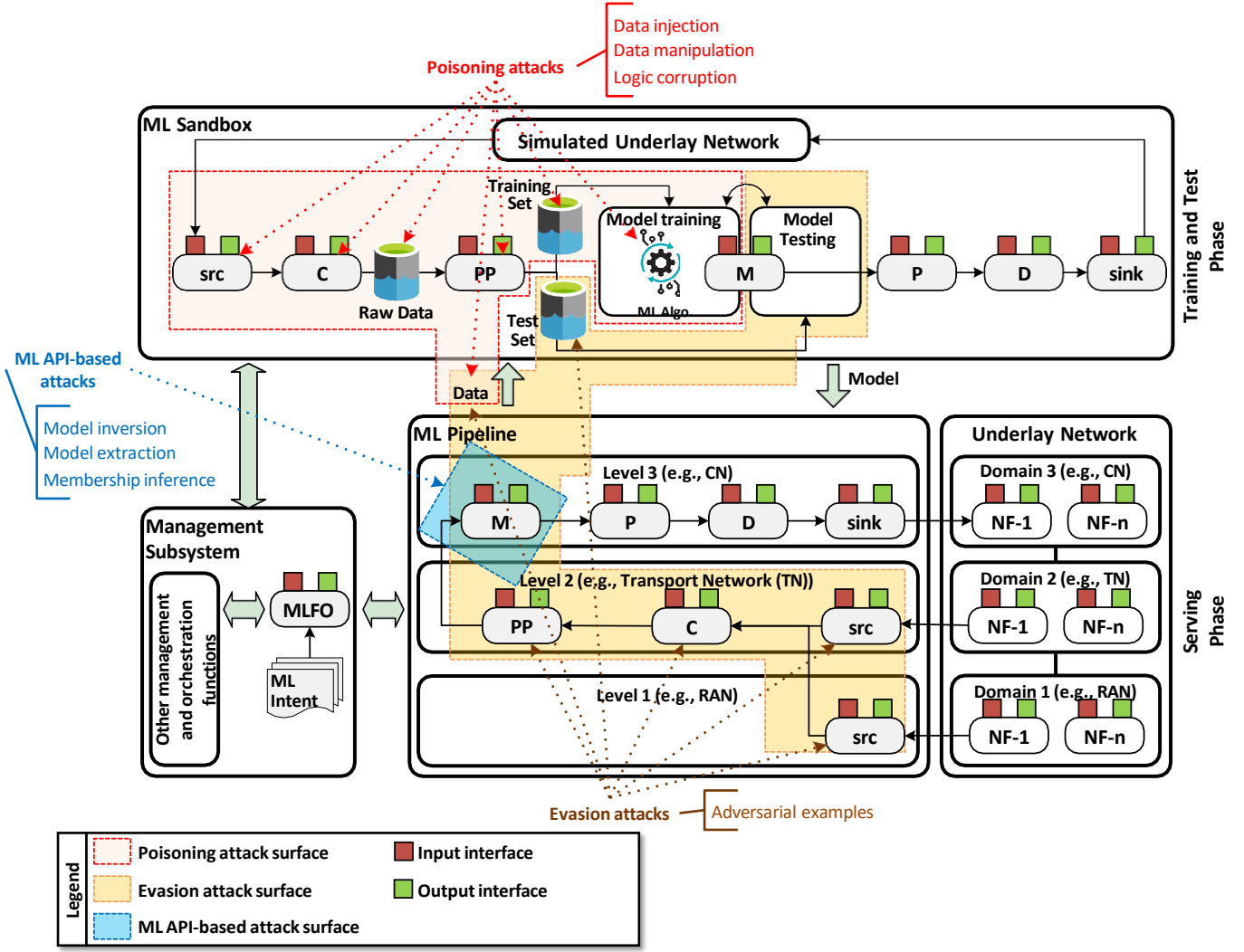

Fig. 5: Mapping of the Adversarial ML Attacks to the ML5G High-Level Architecture.

*2) Defenses Against Evasion Attacks:* A variety of defensive strategies have emerged for defeating evasion attacks, including adversarial training, defensive distillation, ensemble methods, defense Generative Adversarial Networks (GANs), and adversarial concept drift handling techniques. In *adversarial training*, the resilience to evasion attacks is achieved by training the model on a dataset augmented with adversarial examples. *Defensive distillation* is a training strategy that uses the knowledge inferred from a ML model to strengthen its own robustness to adversarial examples. Both adversarial training and defensive distillation implicitly perform gradient masking, which consists in making the model's gradient useless by, for instance, setting it to zero or changing its direction. Indeed, the absence of the real gradient complicates the generation of adversarial examples, allowing the model to exhibit improved robustness. However, this does not prevent that the model may remain vulnerable to adversarial samples crafted using transferability-based black-box attacks. Moreover, it is worth mentioning that the improved robustness brought by adversarial training and defensive distillation comes at the price of a decreased accuracy on clean data. *Ensemble methods* combine multiple models to build a robust model. Ensemble methods have the virtue of improving the model's robustness while increasing its accuracy on clean samples. Nevertheless, the merit of ensemble methods comes at the expense of increased model complexity and computational cost. *Defense GANs* aim to denoise input samples from adversarial perturbations by projecting them on to the range of the GAN's generator before feeding them into the ML model. In other words, they aim to find the closest sample to the adversarial example that the GAN's generator is capable of producing and feed that as an input to the ML model. As the GAN's generator is trained to learn the distribution of the real data, the generated sample will be cleaned from added perturbations. Defense GANs have proven their effectiveness to counter both white-box and black-box attacks. The adversarial perturbations introduced to data result in concept drift; that is, the change in data distribution leading to drop in the ML model performance. Thus, *adversarial concept drift handling techniques*, such as ensemble learning, can be used to face down adversarial attacks by retraining the ML model once a drop in its performance is detected. For instance, an ensemble learning approach tracks the adversarial concept drift by measuring the prediction disagreement between the ensemble models. In fact, an abrupt increase in the prediction disagreement is an indicator of concept drift, that will trigger the retraining of the ensemble models on the new data.

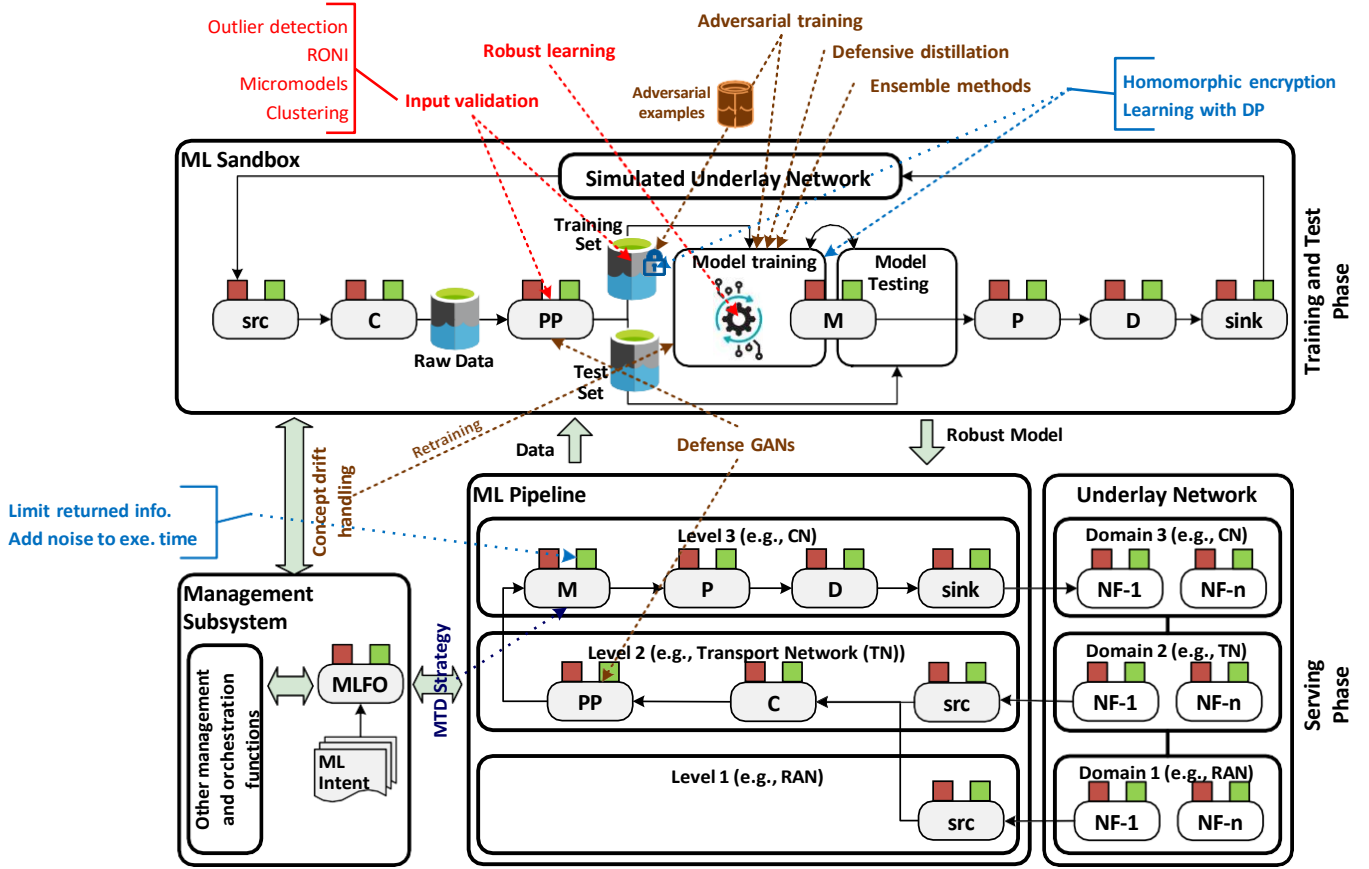

Fig. 6: Mapping of Potential Defenses against ML Attacks to the ML5G High-Level Architecture.

*3) Defenses Against Model's API-based Attacks:* To mitigate ML API-based attacks, various solutions have been proposed, including:

- The *learning with differential privacy* (DP) to prevent the disclosure of training data by making the model prediction independent of an individual input. A differentially private ML model guarantees that its behavior hardly changes when an individual sample is added to or removed from the training dataset. Thus, by looking at the model's output, an adversary cannot ascertain whether an individual input was included in the training dataset or not. To achieve DP, a small, controlled noise is added to the model during its training.
- The use of *homomorphic encryption* which enables model training over encrypted data, thus guaranteeing data privacy. It is worth noting that the major challenge in using this countermeasure is the induced computational complexity.
- The limitation of sensitive information provided by ML APIs by releasing only class labels, filtering out the prediction probabilities of low-probability classes, and rounding the class probabilities. In fact, the danger of revealing the prediction probabilities by the inference API stems from the fact that those probabilities are calculated as a function of the input and the ML model's parameters. Thus, collecting a sufficient number of prediction probabilities and their corresponding inputs, an adversary can easily extract the model's parameters by solving a system of equations where variables are the unknown model's parameters. By hiding the prediction probabilities, revealing only part of them and/or rounding them to a fixed number of decimal places, the adversary is defeated from achieving the goal of building a surrogate model approximating the real one.

- The addition of noise to the execution time of the ML model.

*B. Moving Target Defense*

Given its potential in increasing the attacker's uncertainty, MTD has recently emerged as an effective paradigm in addressing the security concerns of AI, specifically ML techniques. In current practice, a ML model remains static over a long period of time once deployed, which gives the attacker the advantage of time to devise effective adversarial attacks. Thus, introducing dynamicity in a ML system by constantly changing, for instance, the ML algorithm, the features used for training, the model's parameters, helps to improve its robustness. In this vein, Song *et al.* [18] proposed a MTD strategy that dynamically generates new models by retraining independently perturbed versions of the base model after its deployment. To leverage the promising MTD capabilities for thwarting adversarial attacks, a major challenge is to come up with MTD strategies that make the ML model robust without sacrificing its performance and with reduced moving cost. Hence, further research efforts are required in this direction.

*C. Mapping of Potential Defenses to ITU-T's Unified Architecture for ML in 5G and Beyond Networks*

Fig. 6 shows on which components of the ML5G unified architecture the aforementioned defenses could be enforced. The measures to counteract the poisoning attack should be implemented into the ML sandbox subsystem, where the input validation operations and the robust learning may be carried out at the preprocessor node (PP) and model node (M), respectively. To make ML models robust against evasion attacks, proactive countermeasures such as adversarial training, defensive distillation or ensemble methods should be implemented at the model training phase. The retraining operation to handle the malicious concept drift problem could be initiated by the MLFO when the model performance drops considerably. The defense GANs strategy may be applied at the preprocessing stage during the testing and serving phases in order to clean out the input samples from adversarial perturbations. The strategies to tackle model API-based attacks could be incorporated either at the training phase or the deployment phase. It is worth noting that in case of using homomorphic encryption, the data provider should perform preprocessing operations on the data before its encryption, such as removing samples that are redundant or have missing/infinity values, and normalizing the features' values. The MTD strategy to defeat exploratory attacks (i.e., evasion and model API-based attacks) can be established by the MLFO, defining when and how the model move should be made.

V. CONCLUSION

This paper emphasized the key role that AI may play in fostering the security in 5G and beyond networks. Meanwhile, it pointed up the security risks that may come along with the envisioned AI's benefits if their potential or vulnerabilities are leveraged by malicious actors. In view of increasing the resilience to AI threats, we advocated several defense measures while advising on which components of the ML5G unified architecture they could be enforced. Despite the merit of the recommended defenses, each of them has its own limitations and none of them can constitute an all-in-one solution for addressing all AI threats. Thus, a potential research direction is to investigate how those countermeasures could be used in an integrated way to meet both security and performance requirements.

ACKNOWLEDGMENT

This work was supported in part by the European Union's Horizon 2020 research and innovation programme under the INSPIRE-5Gplus project (Grant No. 871808) and MonB5G project (Grant No. 871780); the Academy of Finland Project 6Genesis Flagship (Grant No. 318927); and CSN (Grant No. 311654).

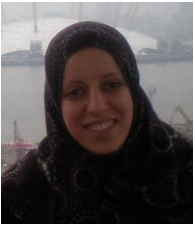
**Chafika Benzaïd** is currently a PostDoc researcher at MOSA!C Lab, Aalto University. She is an associate professor and research fellow in Computer Science Department at University of Sciences and Technology Houari Boumediene (USTHB). She obtained her PhD degree in Computer Sciences from USTHB in 2009. Her current research interests include AI-driven network security and AI security. She serves/served as a TPC member for several international conferences and as a reviewer for multiple international journals.

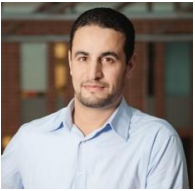
**Tarik Taleb** is Professor at Aalto University and University of Oulu. He is the founder and director of the MOSA!C Lab (www.mosaic-lab.org). Prior to that, he was a senior researcher and 3GPP standards expert at NEC Europe Ltd., Germany. He also worked as assistant professor at Tohoku University, Japan. He received his B.E. degree in information engineering, and his M.Sc. and Ph.D. degrees in information sciences from Tohoku University in 2001, 2003, and 2005, respectively.

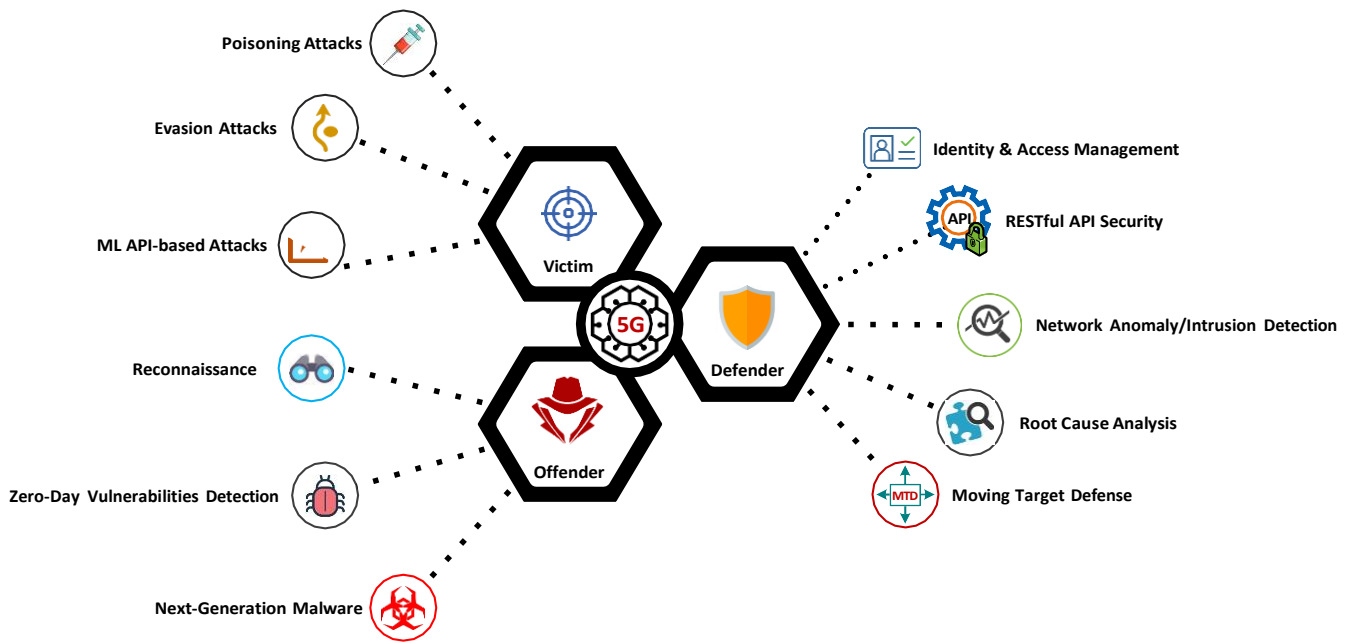

Fig. 1: AI's Impact on Security in 5G and Beyond Networks from its Posture of Defender, Offender or Victim.

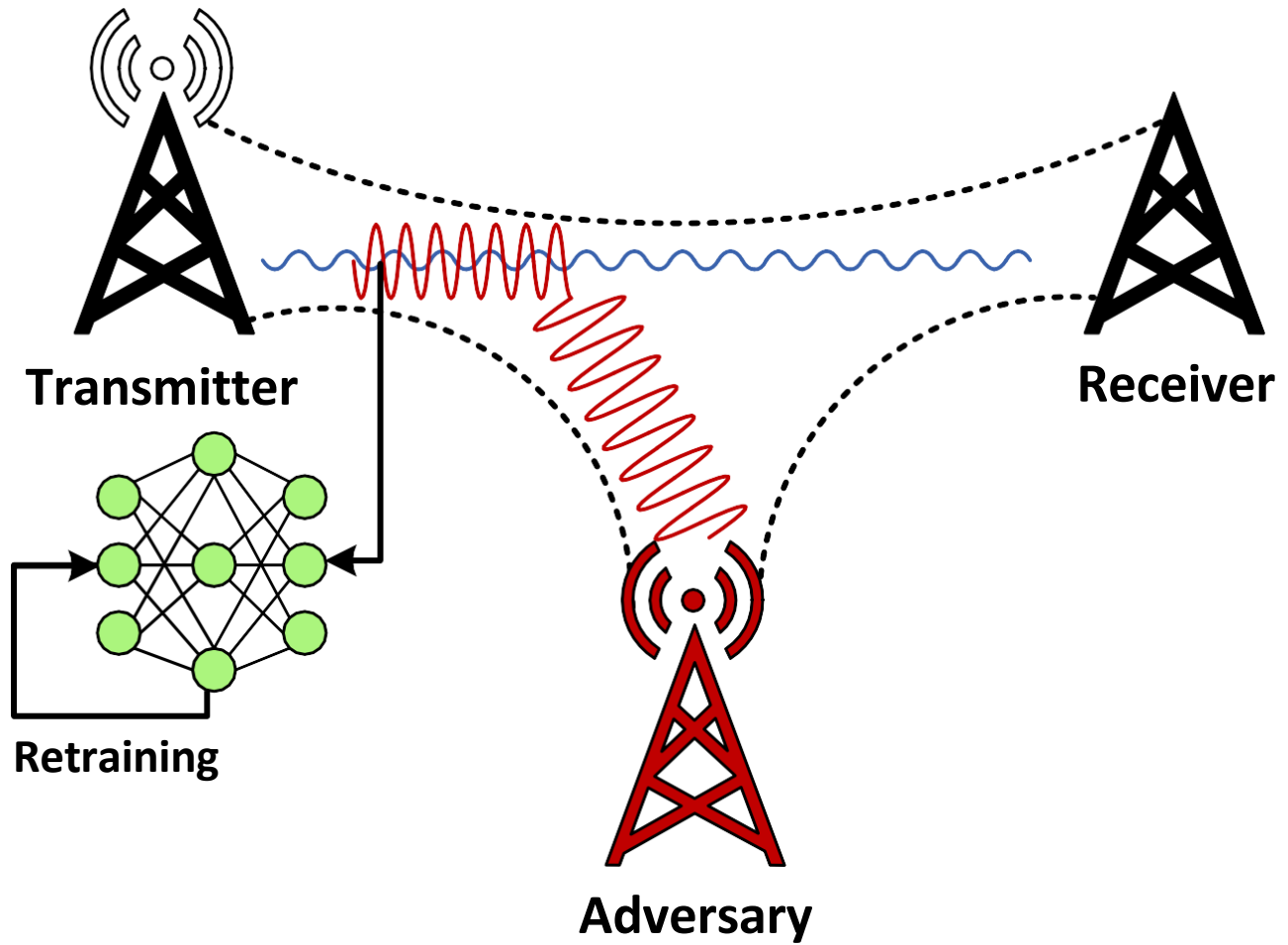

(a) Sensing phase.

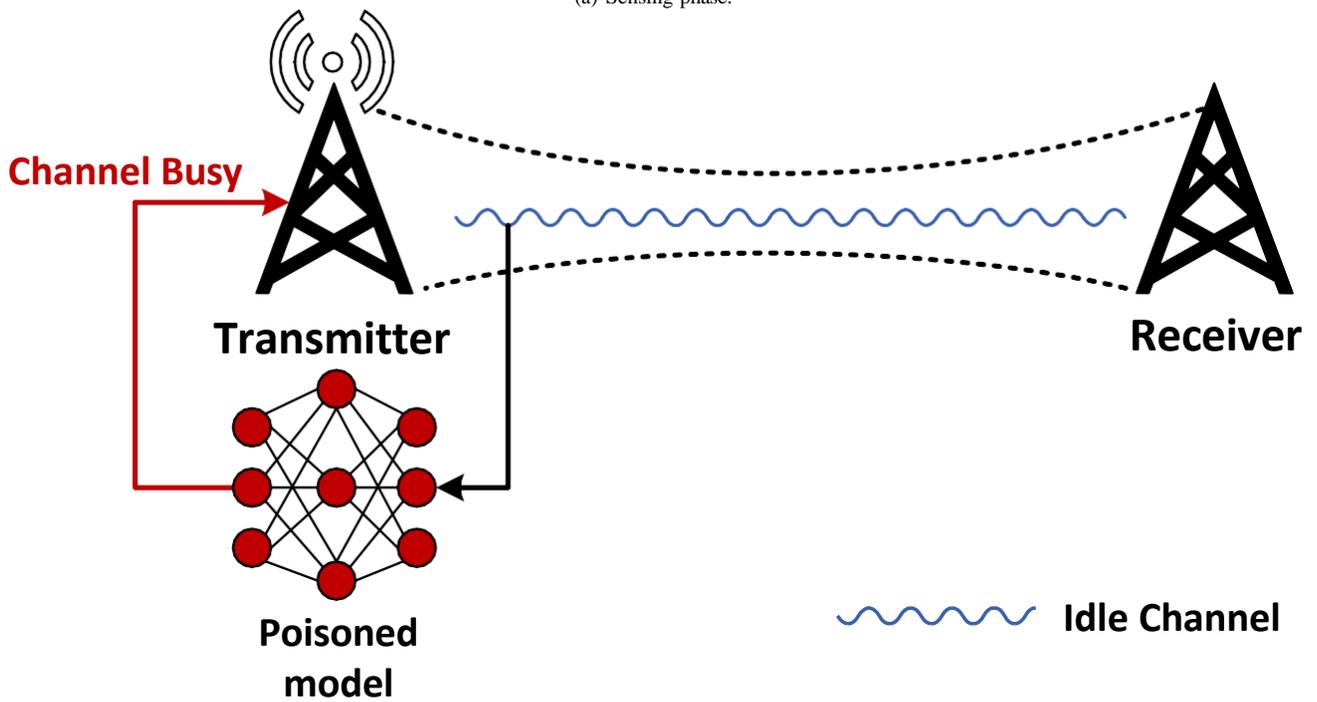

(b) Transmission phase.

Fig. 2: Spectrum Data Poisoning in Cognitive Radio Networks.

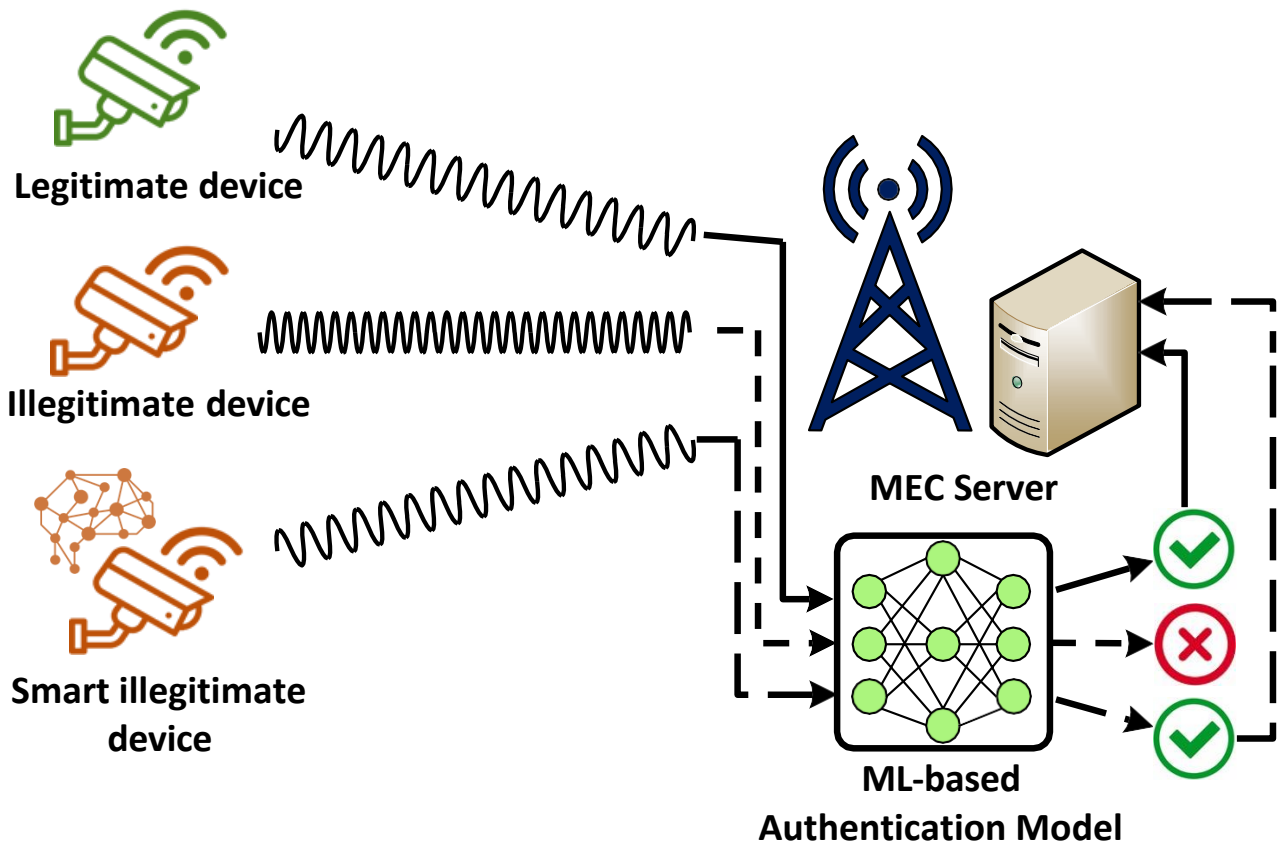

Fig. 3: Adversarial Identity Spoofing.

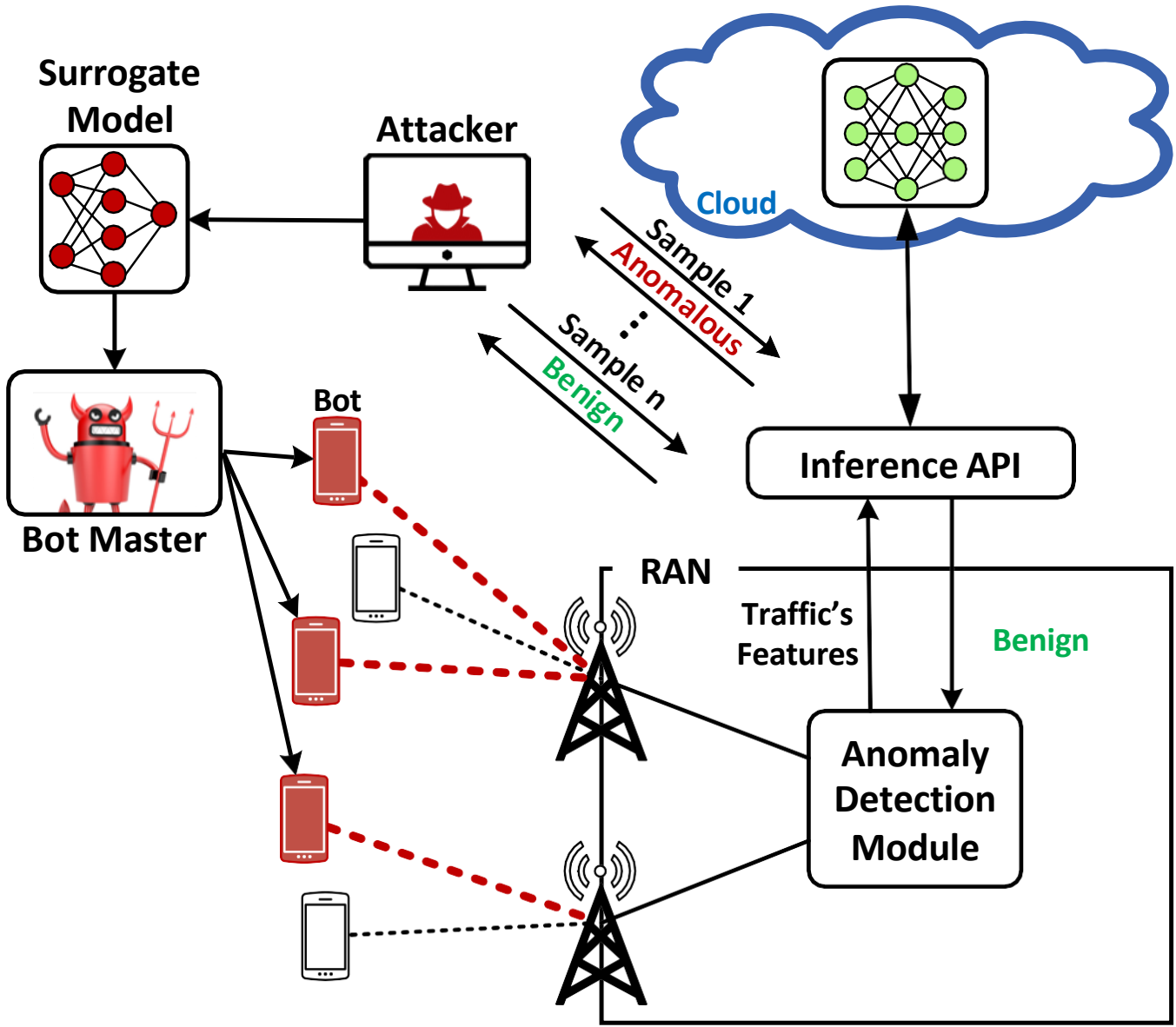

Fig. 4: Model Extraction for Subsequent Evasion Attack Against an AI-based Network Anomaly Detection Module.

14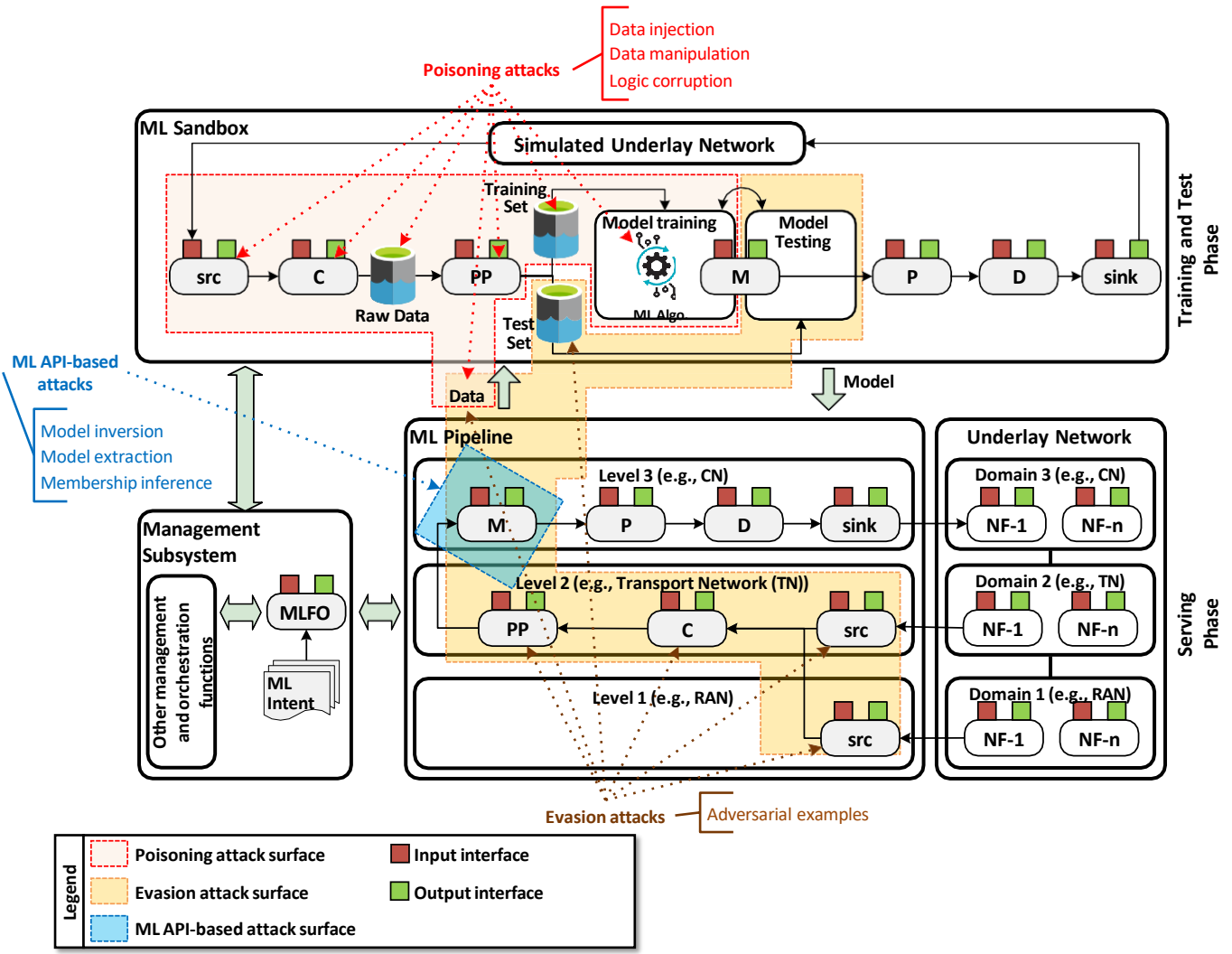

Fig. 5: Mapping of the Adversarial ML Attacks to the ML5G High-Level Architecture.

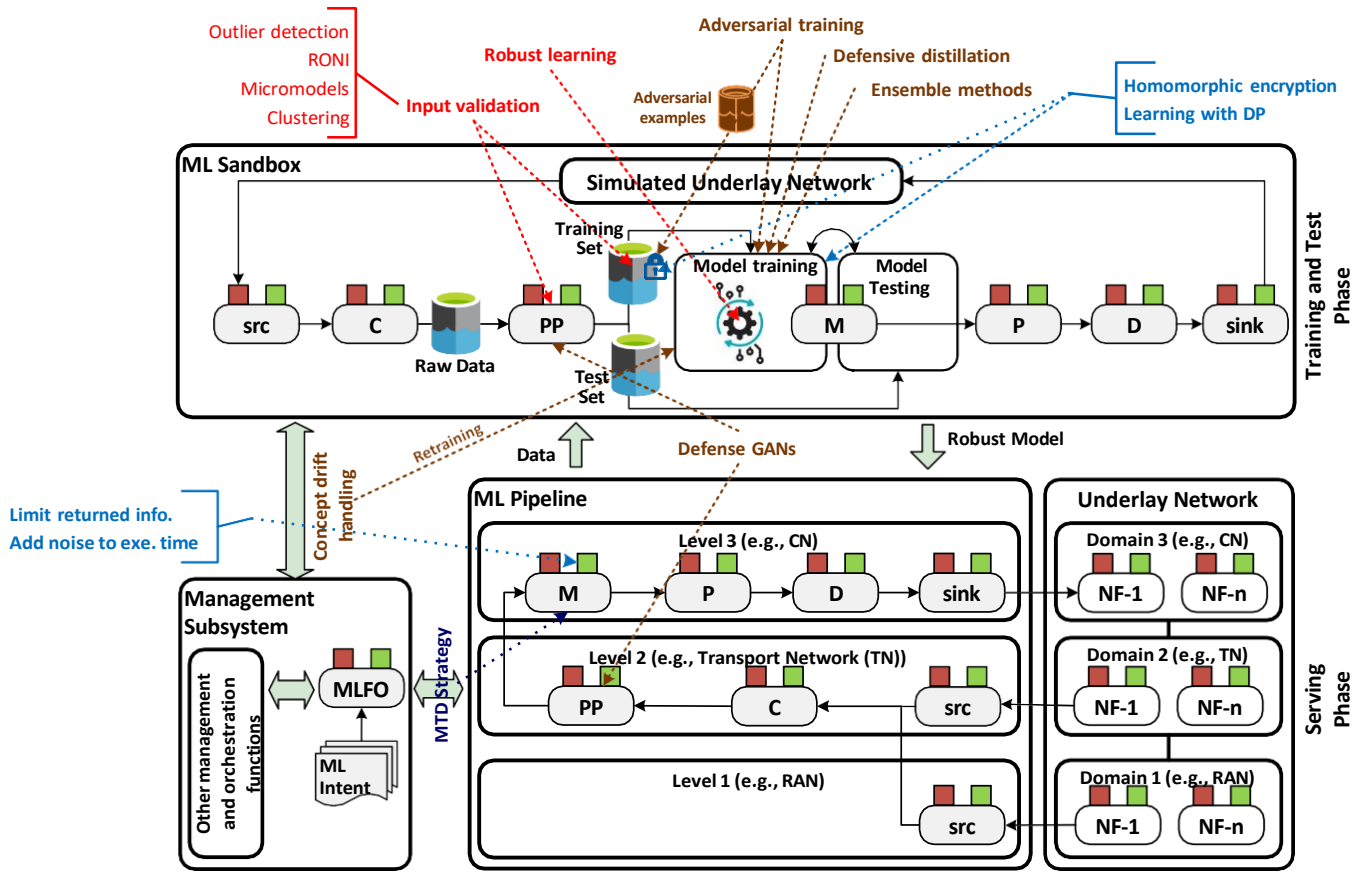

Fig. 6: Mapping of Potential Defenses against ML Attacks to the ML5G High-Level Architecture.